\documentclass[pra,aps,superscriptaddress,floatfix,tightenlines,twocolumn]{revtex4}

\usepackage{amsmath}
\usepackage{amssymb}
\usepackage{amsfonts}
\usepackage{epsfig}
\usepackage{subfigure}
\usepackage[dvips]{color}
\usepackage{bm}
\usepackage{times}
\usepackage{amsthm}

\begin{document}
\title{Quantifying Coherence in Infinite Dimensional Systems}
\author{Yu-Ran Zhang}
\affiliation{Beijing National Laboratory for Condensed Matter Physics, Institute
of Physics, Chinese Academy of Sciences, Beijing 100190, P. R. China}
\author{Lian-He Shao}
\affiliation{College of Computer Science, Shaanxi Normal University, Xi¡¯an 710062, P. R. China}
\affiliation{Beijing National Laboratory for Condensed Matter Physics, Institute
of Physics, Chinese Academy of Sciences, Beijing 100190, P. R. China}
\author{Yongming Li}
\email{liyongm@snnu.edu.cn}
\affiliation{College of Computer Science, Shaanxi Normal University, Xi¡¯an 710062, P. R. China}
\author{Heng Fan}
\email{hfan@iphy.ac.cn}
\affiliation{Beijing National Laboratory for Condensed Matter Physics,
Institute of Physics, Chinese Academy of Sciences, Beijing 100190, P. R. China}
\affiliation{Collaborative Innovation Center of Quantum Matter, Beijing 100190, P. R. China}
\date{\today}
\begin{abstract}
We study the quantification of coherence in infinite dimensional systems, especially the infinite dimensional
bosonic systems in Fock space.
We show that given the energy constraints, the
relative entropy of coherence serves as a well-defined quantification of coherence in infinite dimensional
systems. Via using the relative entropy of coherence, we also generalize the problem to multi-mode Fock
space and special examples are considered.  It is shown that with a finite average particle number, increasing
the number of modes of light can enhance the relative entropy of coherence. With the mean energy constraint,
our results can also be extended to other infinite-dimensional systems.
\end{abstract}
\maketitle

Quantum coherence arising from quantum superposition principle is a fundamental
aspect of quantum physics \cite{NC}. The laser \cite{laser} and superfluidity \cite{sf}
are examples of quantum coherence, whose effects are evident at the macroscopic scale.
However, the framework of quantification of coherence has only been methodically
investigated recently. The first attempt to address the classification of quantum coherence as physical
resources by T. Baumgratz $et.$ $al.$, who have established a rigorous framework for the
quantification of coherence based on distance measures in finite dimensional setting~\cite{coherence}.
With such a fundational framework for coherence, one can find the appropriate distance measures to
quantify coherence in a fixed basis by measuring the distance between the quantum state
$\hat{\rho}$ and its nearest incoherent state. After the framework for coherence has been proposed,
it receives increasing attentions. A. Streltsov $et.$ $al.$ have used entanglement to quantify quantum
coherence, which provides the operational quantification of coherence~\cite{Streltsov15}. S. Du $et.$ $al.$
focused on the interconversion of coherent states by means of incoherent operations using the
concept of majorization relations~\cite{SDU13}. Z. Xi $et.$ $al.$ have given a clear quantitative analysis and
operational connections between relative entropy of coherence, quantum discord and one-way
quantum deficit in the bipartite quantum system~\cite{ZJXI14}. T. Bromley $et.$ $al.$ have found
freezing conditions in which coherence remains unchanged during the nonunitary dynamics~\cite{Bromley14}.
Up to now, all the results for quantifying the quantum coherence are assumed the finite dimensional
setting, which is neither necessary nor desirable. In consideration of the relevant physical situations
 such as quantum optics states of light, it must require further investigations on
infinite dimensional systems.

In this paper, we aim to investigate the quantification of coherence in infinite dimensional
systems. Specificly, we focus on the infinite dimensional bosonic systems in Fock space \cite{Fock}
which are used to describe the most notable quantum optics states of light \cite{qo} and Gaussian states
\cite{Gaussian1,Gaussian2,Gaussian3}. We show that when considering the energy constraints, the
relative entropy of coherence serves as a well-defined quantification of coherence in infinite dimensional
systems and the $l_1$ norm of coherence fails. Via using the relative entropy of coherence, we also
generalize the results to multi-mode Fock space and special examples are considered.  It is shown that
with a finite average particle number, increasing the number of modes of light can enhance the relative
entropy of coherence. Our results can also be extended to other infinite-dimensional systems with energy
constraints. Our work investigates special and experimentally relevant cases and the most general and easy
to use quantifiers, which is  significant and essential in quantum physics as well as quantum optics.

Given the postulates presented in Ref.~\cite{coherence},
any proper measure of the coherence $C(\hat{\rho})$ must satisfy the following conditions:

$(C1)$ $C(\hat{\rho})\geq0$ for $\forall$ $\hat{\rho}\in\mathcal{H}$ and $C(\hat{\delta})=0$
\emph{iff} $\forall\hat{\delta}\in\mathcal{I}$.

$(C2a)$ Monotonicity under all the incoherent completely positive and trace-preserving (ICPTP) maps:
$C(\hat{\rho})\geq C(\Phi_{\textrm{ICPTP}}(\hat{\rho}))$, where $\Phi_{\textrm{ICPTP}}(\hat{\rho})=\sum_{n}\hat{K}_{n}^{\dag}\hat{\rho}\hat{K}_n$
and $\{\hat{K}_n\}$ is a set of Kraus operators that satisfies $\sum\hat{K}_{n}^{\dag}\hat{K}_n=\mathbb{I}$ and
$\hat{K}_{n}\mathcal{I}\hat{K}_n\subset \mathcal{I}$.

$(C2b)$ Monotonicity for average coherence under subselection based on measurement outcomes:
$C(\hat{\rho}) \geq\sum_{n}p_n C(\hat{\rho}_n)$,
where $\hat{\rho}_n = {\hat{K}_n\hat{\rho}\hat{K}^{\dag}_n}/{p_n}$
and $p_n = Tr(\hat{K}_n\hat{\rho}K^{\dag}_{n})
$ for all $\{\hat{K}_n\}$ with$\sum_{n}\hat{K}_{n}^{\dag}\hat{K}_n=\mathbb{I}$ and
$\hat{K}_{n}\mathcal{I}\hat{K}_n\subset \mathcal{I}$.

$(C3)$ Nonincreasing under the mixing of quantum states: $\sum_{n}p_nC(\hat{\rho}_{n})\geq C(\sum_n p_n \hat{\rho}_n)$.

Two kinds of measures for coherence in finite dimensional systems \cite{coherence} satisfy
all the conditions mentioned above include: the relative entropy of coherence defined as
\begin{eqnarray}
C_{\textrm{rel.ent.}}(\hat{\rho})=S(\hat{\rho}_{\textrm{diag}})-S(\hat{\rho})\label{c1}
\end{eqnarray}
and the $l_1$ norm of coherence defined as
\begin{eqnarray}
C_{l_1}(\hat{\rho})=\sum_{i\neq j}|\rho_{ij}|\label{c2}
\end{eqnarray}
where $\hat{\rho}=\sum_{ij}\rho_{ij}|i\rangle\langle j|$ and $\hat{\rho}_{\textrm{diag}}=\sum_{i}\rho_{ii}|i\rangle\langle i|$.
It has been shown that the promising fidelity of coherence does not in general satisfy $(C2b)$
under the subselection of the measurement condition \cite{shao}.

\begin{figure*}[t]
 \centering
\includegraphics[width=0.93\textwidth]{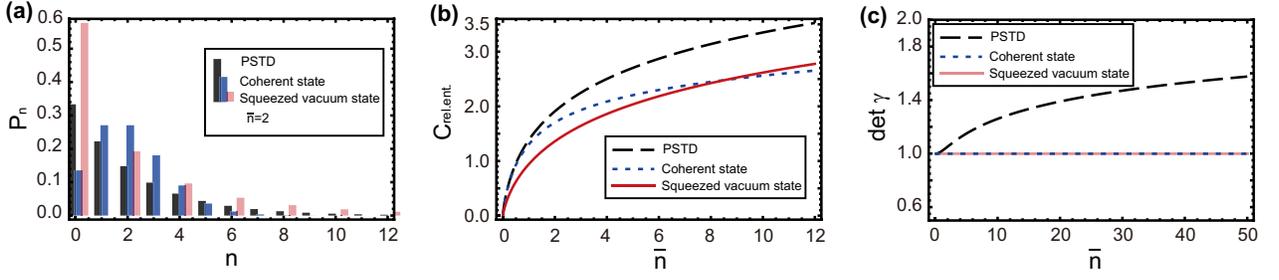}\\
\caption{(color online) (a) Photon number distributions of PSTD, coherent state and squeezed vacuum state
against average particle number. (b) Relative entropies of coherence of PSTD, coherent state and squeezed vacuum state against average particle number. (c) Determinants of the coherence variances matrices $\gamma$ of these
three states against the mean particle number.}\label{f1}
\end{figure*}

Generally, the bosonic single mode Hilbert space $\mathcal{H}$
is spanned by an uncountable basis $\{|n\rangle\}_{n=0}^{\infty}$ called the Fock (number state) basis.
Fock states are the eigenstates of the number operator $\hat{n}:=\hat{a}^{\dag}\hat{a}$ where we have
$\hat{a}|n\rangle=\sqrt{n}|n-1\rangle$ and $\hat{a}^{\dag}|n\rangle=\sqrt{n+1}|n+1\rangle$.
Referring to development of entanglement theory in infinite dimensional systems, the problem of
quantification of coherence can addressed by requiring energy constraints \cite{infinite}, which is
experimentally relevant. Here and after, we require a new condition for this case ($C4$): if the first order moment,
the average particle number, is finite $\langle\hat{n}\rangle<\infty$, it should fulfill $C(\hat{\rho})<\infty$.

Given the proper definition of incoherent states, incoherent operations and
maximal coherent states, the proofs of these two definitions do not require the
finite dimensional setting as there are very relevant physical situations that
require infinite dimensional systems for their description.
The incoherent states and incoherent operations defined in Ref.~\cite{coherence} can be
easily generalized to the case in infinite dimensional systems. In the Fock space, the  set
of incoherent state can be defined as $\mathcal{I}\subset\mathcal{H}$ and all density
operators $ \hat{\delta}\in\mathcal{I}$ are of the form
$\hat{\delta}=\sum_{n=0}^{\infty}\delta_{n}|n\rangle\langle n|$.
For $(C2)$, Kraus operators $\{\hat{K}_n\}$  satisfying $\sum\hat{K}_{n}^{\dag}\hat{K}_n=\mathbb{I}$
and
$\hat{K}_{n}\mathcal{I}\hat{K}_n\subset \mathcal{I}$ are $d_n\times d_\textrm{in}$ matrices where $d_\textrm{in}\rightarrow\infty$.
Given these premises, our problem turns to be
verifying condition $(C4)$: whether these quantifications of coherence fulfilling $(C1$-$3)$ can
serve as a unit for coherence or be finite $C(\rho)<\infty$ when the energy constraint is
taken into consideration. That is, incoherent states, maximal coherent states and the maximum quantification
of coherence should be well-defined.


At first, we show that relative entropy of coherence $C_{\textrm{rel.ent.}}$ fulfills the
requirements of quantification of coherence for the states in the infinite dimensional
Hilbert space. At the beginning, we show that diagonal mixed states such as thermal states have zero coherence
$C_{\textrm{rel.ent.}}=0$.
When mean particle number is finite, we can figure out the maximal coherent
state as
\begin{eqnarray}
|\psi_{m}\rangle=\sum_{n=0}^{\infty}\frac{\bar{n}^{n/2}}{(\bar{n}+1)^{(n+1)/2}}e^{i\varphi_n}|n\rangle\label{os}
\end{eqnarray}
which makes $(C4)$ saturated:
\begin{eqnarray}
C_\textrm{rel.ent.}^\textrm{max}=(\bar{n}+1)\log(\bar{n}+1)-\bar{n}\log\bar{n}<\infty.
\end{eqnarray}
This result can be directly obtained from the fact that the thermal state as
$\hat{\rho}^{\textrm{th}}(\bar{n})=\sum_{n=0}^{\infty}({\bar{n}^{n}}/{(\bar{n}+1)^{n+1}})|n\rangle\langle n|$
reaches the maximum von Neumann entropy given a fixed average particle number $\bar{n}:=\langle\hat{n}\rangle$.
The normalized second-order correlation function can be calculated as
$g^{2}(0)={\langle\hat{a}^{\dag}\hat{a}^{\dag}\hat{a}\hat{a}\rangle}/{\langle \hat{n}\rangle^2}=2$
which is the same as the thermal state.
Given a linear phase generation $\varphi_n=n\varphi$, the state (\ref{os}), a pure state with a thermal distribution
(PSTD), has been shown to be the eigenstate of the SG-phase operator $\sum_{n=0}^{\infty}|n\rangle\langle n+1|$
with eigenvalue $\sqrt{\bar{n}/(\bar{n}+1)}e^{i\varphi}$ \cite{pstd}. A proposal of the generation of PSTD
in the ``particle box'' \cite{photonbox} has been also presented in Ref.~\cite{pstd}. Compared with two well-known
Gaussian states, coherent state $|\alpha\rangle:=\hat{D}(\alpha)|0\rangle$ and squeezed vacuum state
$|0,\xi\rangle=\hat{S}(\xi)|0\rangle$, the particle number distribution and coherence quantification of
relative entropy are shown in Fig.~\ref{f1}(a) and \ref{f1}(b), respectively. In Fig.~\ref{f1}(c), the determinants
of the coherence variances matrices $\gamma$ of these three states against the mean particle number are given.
Since a Gaussian state is pure \emph{iff} $\det \gamma = 1$ \cite{Gaussian1,Gaussian2,Gaussian3}, we conclude
that PSTD with form (\ref{os}) is a non-Gaussian state, except for $\bar{n}\rightarrow0$. Therefore, PSTD
can not be easily constructed by squeezing and displacement operator on vacuum state.
For details, please see APPENDIX. Therefore,  we conclude that relative entropy of coherence is an appropriate
quantification of coherence even  in infinite dimensional systems.

Next, given a fixed average particle number in Fock space, we show that no maximal coherent state can be
found to maximize the $l_1$ norm of coherence. With a set of particle number distributions
$\{P_n\}$ of a pure state, the identity condition: $\sum_{n=0}^{\infty}P_n=1$ and the finite energy constraint
$(C4)$: $\sum_{n=0}^{\infty}nP_n=\bar{n}<\infty$ should be two  constraint conditions. Obviously, for any mixed state,
we can find a pure state with larger $l1_1$ norm of
coherence. Then, $l_1$ norm of coherence of a pure state can be written as
\begin{eqnarray}
C_{l_1}(\hat{\rho})=\sum_{m,n=0}^{\infty}\sqrt{P_mP_n}-1.\label{cl1}
\end{eqnarray}
The maximum of $l_1$ norm of coherence should occur as the first variation is zero
$\delta C_{l_1}=\sum_{m,n=0}^{\infty}\sqrt{{P_{m}}/{P_{n}}}\delta P_{n}=0$.
Via using the method of Lagrange multipliers with two Lagrange multiplier $\lambda_1$ and $\lambda_2$,
a series of equations can be obtained:
\begin{eqnarray}
\frac{\sum_{m=0}^{\infty}\sqrt{P_{m}}}{\sqrt{P_{n}}}+\lambda_1 n+\lambda_2=0
\end{eqnarray}
the solutions of which obviously do not satisfy $C(4)$. Moreover, $\sum_{m=0}^{\infty}\sqrt{P_{m}}$
that relates to the Riemann Zeta function \cite{JPA} is infinite. Mathematically, $l_1$ norm of coherence (\ref{cl1})
is a concave function in probability space which makes the Karush-Kuhn-Tucker (KKT) conditions \cite{KKT} also
sufficient for the optimality. However, no proper solutions can be figured out such that this optimal problem
may not be derived analytically. Therefore, the $l_1$ norm of coherence does not seem to be a well-defined
quantification of coherence in Fock space because it does not have a well-defined maximal coherent state such that
the quantification is finite. We here note that with a stronger condition ($C4'$): the second order moment is finite
$\langle\hat{n}^2\rangle<\infty$, we can find a well-defined maximal coherent state for the $l_1$ norm of coherence
and $C(\hat{\rho})<\infty$ could be met.

\begin{figure}[t]
 \centering
\includegraphics[width=0.3\textwidth]{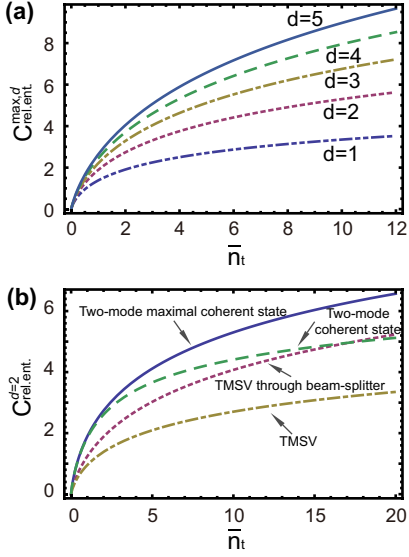}\\
\caption{(color online) Relative entropies of coherence of multi-mode states. (a) Relative entropies of
coherence of maximal coherent states with $d=1,2\cdots,5$. (b) For two-mode states $d=2$, relative
entropies of coherence of maximal coherent state, two-mode coherent state, TMSV and TMSV through
a  50:50 beam-splitter.}\label{f2}
\end{figure}

Since we have shown that the relative entropy of coherence $C_{\textrm{rel.ent.}}$ fulfills the
requirements of quantification of coherence even for the states in single mode Fock space, we then generalize
this result to $d$-mode Fock space $\mathcal{H}=\otimes_{i=1}^{d}\mathcal{H}_{i}$. It has
an uncountable basis $\{\otimes_{i=1}^{d}|n_{i}\rangle_i\}$ and probability distributions $\{P_{\bm{n}}\}$
where the vector is defined as $\bm{n}=(n_1,n_2\cdots,n_d)$ and we define $|\bm{n}|_{1}=\sum_{i=1}^{d}n_{i}$.
After simple calculations, the maximal coherent state should has a distribution as
$P_{\bm{n}}^\textrm{max}={\bar{n}_\textrm{t}^{|\bm{n}|_1}}/[{(\bar{n}_\textrm{t}+1)^{|\bm{n}|_1+1}}\mathbb{C}_{|\bm{n}|_1+d-1}^{d-1}]$
with finite average total particle number defined as $\bar{n}_\textrm{t}:=\sum_{\bm{n}}P_{\bm{n}}|\bm{n}|_1$. The maximum
relative entropy of coherence for $d$-mode Fock space can be calculated as
\begin{equation}
C_\textrm{rel.ent.}^{\textrm{max},d}=C_\textrm{rel.ent.}^{\textrm{max},d=1}+S_d(\bar{n}_\textrm{t})
\end{equation}
where $S_d(\bar{n}_\textrm{t}):=\sum_{n=0}^{\infty}({\bar{n}^{n}_\textrm{t}}/{(\bar{n}_\textrm{t}+1)^{n+1}})\log(\mathbb{C}_{n+d-1}^{d-1})$ is a convergent series.
Since $S_d(\bar{n}_\textrm{t})>S_{d'}(\bar{n}_\textrm{t})$ if $d>d'$, we show in Fig.~\ref{f2}(a) that given a fixed
average total particle number $\bar{n}_\textrm{t}$, relative entropy of coherence increases as the the number of modes $d$ increases. This result is significant that with a finite average particle number increasing the
number of modes of light can enhance the coherence as a resource in quantum information processing. The advantages of multimode quantum optics have been recently interpreted in quantum metrology \cite{qm}.

We then consider two-mode coherent state $|\alpha\rangle_{1}|\alpha\rangle_{2}$, two-mode squeezed
vacuum (TMSV) state and TMSV passing a 50:50 beam-splitter as special examples. The last case has been
shown to be efficient to beat the shot noise limit (SNL) in the quantum metrology \cite{dowling}. TMSV can
be written as $|\textrm{TMSV}\rangle=\sum_{n=0}^{\infty}{(\bar{n}_{\textrm{t}}/2)^{\frac{n}{2}}/(\bar{n}_{\textrm{t}}/2+1)^{\frac{n+1}{2}}}|n\rangle_1|n\rangle_2$ and a TMSV through
a 50:50 beam-splitter is written as \cite{dowling,JPA}
\begin{widetext}
\begin{eqnarray}
\hat{U}_{\textrm{BS}}|\textrm{TMSV}\rangle=\sum_{n=0}^{\infty}\frac{(\frac{\bar{n}_{\textrm{t}}}{2})^{\frac{n}{2}}}{{(\frac{\bar{n}_{\textrm{t}}}{2}+1)^{\frac{n+1}{2}}}}
\sum_{k=0}^{n}(-1)^k\frac{\mathbb{C}_{n}^{k}[(2n-2k)!(2k)!]^{\frac{1}{2}}}{2^{{n}}n!}|2n-2k\rangle_1|2k\rangle_2,
\end{eqnarray}
\end{widetext}
where $\hat{U}_{\textrm{BS}}:=\exp[i\pi(\hat{a}^{\dag}\hat{b}+\hat{a}\hat{b}^{\dag})/2]$ is the unitary
transformation of a 50:50 beam-splitter with $\hat{a}$ ($\hat{a}^{\dag}$) and $\hat{b}$ ($\hat{b}^{\dag}$)
the annihilation (creation) operators for two modes, respectively. With the maximal coherent states for $d=2$,
we show in Fig.~\ref{f2}(b) the relative entropies of coherence of these three Gaussian two-mode states
against the total average particle number. It is obvious that TMSV through a 50:50 beam-splitter has a larger
coherence than TMSV.


In conclusion, we investigate the quantification of coherence in infinite dimensional systems, since
there are very relevant physical situations that require infinite dimensional systems for their description.
A new constraint condition $(C4)$ is suggested for this problem, with which
the relative entropy of coherence is shown to be  a well-defined quantification of coherence in
infinite dimensional systems but the $l_1$ norm of coherence fails. We also consider quantifying
coherence in the multi-mode Fock space. Given a fixed average total particle number, relative entropy
of coherence increases as the the
number of modes increases, which is significant that the coherence as a resource in quantum
information processing is larger when increasing the number of modes. This work investigates
experimentally relevant infinite dimensional systems and the most general and easy to use quantifiers,
which is important  for experimental and theoretic applications in quantum physics as well as quantum
optics. Moreover, our results can be easily extended to other infinite-dimensional systems.

We thank Si-Yuan Liu for the valuable discussions. This work was supported the
NSFC under grant No. 11175248, the grant from Chinese
Academy of Sciences (XDB01010000).

\appendix
\section{Relative entropy of coherence of coherent state and squeezed vacuum state}
The well-known coherent state can be written as
$|\alpha\rangle=e^{-|\alpha|^2/2}\sum_{n=0}^{\infty}{\alpha^n}/{\sqrt{n!}}|n\rangle$
with a  particle number distribution $P^\textrm{cs}_n=e^{-\bar{n}}{\bar{n}^n}/{n!}$ and $\bar{n}=|\alpha|^2$.
The relative entropy as a quantification of coherence can be calculated as
\begin{eqnarray}
C^\textrm{cs}_{\textrm{rel.ent.}}=
e^{-\bar{n}}\sum_{n=0}^{\infty}\frac{\bar{n}^n\log n!}{n!}-\bar{n}\log\frac{\bar{n}}{e},
\end{eqnarray}
which is shown in Fig.~\ref{f1}(b).
A squeezed state $|\alpha,\xi\rangle$ may be generated by first acting with
the squeeze operator $\hat{S}(\xi)$ on the vacuum followed by the
displacement operator $\hat{D}(\alpha)$
with particle number distribution ($\xi=re^{i2\phi}$) \cite{squeeze}
\begin{eqnarray}
P_n^\textrm{ss}&=&\frac{\exp\left[-|\alpha|^2-\frac{1}{2}\tanh r\left(\alpha^{*2}e^{i\phi}+\alpha^2e^{-i\phi}\right)\right]}{2^nn!\cosh r}\nonumber\\
&\times&\tanh^n r\left|H_{n}\left(\frac{\alpha+\alpha^* e^{i\phi}\tanh r}{\sqrt{2e^{i\phi}\tanh r}}\right)\right|^2,
\end{eqnarray}
where $H_{n}(z)$ is the $n$th Hermite polynomial.
For squeezed vacuum state, $\alpha=0$ and $|H_{n}(0)|=2^{n/2}(n-1)!!$ when $n$ is even, we obtain that
\begin{eqnarray}
P_n^\textrm{sv}=\frac{\tanh^n r[(n-1)!!]^2}{n!\cosh r},
\end{eqnarray}
where $\bar{n}=\sinh^2 r$. Then we can calculate the relative entropy
using $C^\textrm{sv}_{\textrm{rel.ent.}}=\sum_{n=0}^{\infty}P^\textrm{sv}_n \log P^\textrm{sv}_n$ in Fig.~\ref{f1}(b).

\section{Covariance matrix of PSTD}
Canonical variables can be written in terms of creation and annihilation operators
\begin{eqnarray}
\hat{x}=\frac{1}{\sqrt{2}}(\hat{a}+\hat{a}^{\dag}),\ \hat{p}=\frac{1}{\sqrt{2}i}(\hat{a}-\hat{a}^{\dag}),\ \hbar=1.
\end{eqnarray}
For the one-mode state, by defining a vectorial operator $\bm{R}=(\hat{x},\hat{p})$, we can calculate
the covariance matrix 
\begin{eqnarray}
\gamma&=&2\left(\begin{array}{c c}\textrm{Cov}_{\rho}(\hat{x},\hat{x})&\textrm{Cov}_{\rho}(\hat{x},\hat{p})\\\textrm{Cov}_{\rho}(\hat{p},\hat{x})&\textrm{Cov}_{\rho}(\hat{p},\hat{p})\end{array}\right)-iJ_{1}\\
&=&2\left(\begin{array}{c c}\bar{n}+\frac{1}{2}+\langle\hat{a}^2\rangle-2\langle\hat{a}\rangle^2&0\\0&\bar{n}+\frac{1}{2}-\langle\hat{a}^2\rangle\end{array}\right).\label{cm}
\end{eqnarray}
where $J_{1}=(_{-1}^{\ 0}{\ }^{1}_{0} )$ and
\begin{eqnarray}
\langle\hat{a}^2\rangle&=&\frac{\bar{n}}{(\bar{n}+1)^2}\sum_{n=0}^{\infty}\left(\frac{\bar{n}}{\bar{n}+1}\right)^n\sqrt{n+2}\sqrt{n+1}\\
\langle\hat{a}\rangle
&=&{\textrm{Li}_{-\frac{1}{2}}\left(\frac{\bar{n}}{\bar{n}+1}\right)}/{\sqrt{\bar{n}(\bar{n}+1)}}
\end{eqnarray}
with $\textrm{Li}_{k}(z)=\sum_{n=1}^{\infty}z^n/n^k$ the polylogarithm function. The determinant of
the covariance matrix (\ref{cm}) is calculated numerically and is shown in Fig.~\ref{f1}(c) against the
mean particle number.

\end{document}